\documentclass[a4paper,11pt]{article}
\usepackage{pos}

\outer\def\gtae {$\buildrel {\lower3pt\hbox{$>$}} \over 
{\lower2pt\hbox{$\sim$}} $}
\outer\def\ltae {$\buildrel {\lower3pt\hbox{$<$}} \over 
{\lower2pt\hbox{$\sim$}} $}

\title{Outbursts in ultra-compact AM CVn binaries}
\ShortTitle{AM CVn binaries}

\author{Gavin Ramsay}

\affiliation{Armagh Observatory \& Planetarium,
  College Hill, Armagh, BT61 9DG, N Ireland, UK}

\emailAdd{gavin.ramsay@armagh.ac.uk}

\abstract{AM CVn binaries are the most compact of accreting binaries
  having orbital periods in the range $\sim$5-70 min. They consist of
  a white dwarf accreting hydrogen deficient material from a
  degenerate or semi-degenerate star and are predicted to be amongst
  the verification sources for future gravitational wave observatories
  such as LISA. Using the recent catalogue of Green et al (2025) I
  focus attention on the orbital period range in which outbursts are
  seen from AM CVn's. I examine in more detail the outburst properties
  of KL Dra which has an outburst every few months and has many
  sectors of {\sl TESS} data as an open resource. Using observational
  data on the outbursting systems in general, I compare the outburst
  recurrence time, duration and amplitude as a function of orbital
  period with the predictions of the disc instability model. The
  recurrence time is well described, although there is some evidence
  that the amount of material in the disc at the end of the quiescence
  phase is less than earlier model assumptions. The distribution of
  the outburst duration appears to be dependent on the cadence of the
  observations and how it is defined. Similarly the amplitude
  distribution is dependent on cadence and the filter, which causes an
  apparent spread in distribution. Both of these features need to be
  systematically studied using consistent benchmarks. AM CVn binaries
  remain an excellent sources to test models which aim to predict the
  properties of disc accreting systems.}

\FullConference{87th Fujihara Seminar: The 50th Anniversary Workshop
  of the Disk Instability Model in Compact Binary Stars
  (DIM50TH2025), 22-26 September 2025, Tomakomai, Japan\\}

\begin{document}

\maketitle

\section{Introduction}

AM CVn binaries are the most compact of accreting binaries (orbital
period, $P_{\rm orb}$, \ltae 70 min) and consist of a white dwarf
accreting material from another white dwarf or semi-degenerate
star. The archetype, AM CVn, was identified as having broad double
helium absorption lines in the 1950's, but it was not until high speed
photometry, which showed a modulation on a period of $\sim$18 min,
that it was revealed as a possible very compact binary system
(\citet{Smak1967}).  However, it was only decades later that
spectroscopy revealed a $P_{\rm orb}$ of 17.15 min (\citet{Nelemans2001}), with
a photometric period of 17.52 min being the superhump period which
is the signature of a precessing accretion disc.

These ultra compact accreting binaries are important for the following
reasons. They have negligible amounts of helium present in their
spectrum -- they are therefore excellent systems to compare and
contrast helium dominated accretion flows with the well studied
hydrogen dominated accretion flows seen in cataclysmic variables
(CVs). Their compact nature implies they should be strong sources of
persistent gravitational waves. Indeed, they are the verfication
sources for ESA's Lisa gravitational wave constellation due to be
launched in the mid 2030's (see \citet{Kupfer2024} and Simone
Scaringi's talk at this workshop). Thirdly, their space density is a
sensitive test for population synthesis models which predict their
number. For a full review of AM CVn binaries in general see
\citet{Solheim2010}

This short overview is focussed on how the long term behaviour of AM
CVn's can be used to test the theoretical models which predict their
observational properties. For many years the thermal-viscous disk
instability scenario (e.g. \citet{MeyerMeyerH1981}) has been modelled
using the Disc Instability Model (DIM) which is the subject of this
workshop. Whilst it has been successful in explaining many of the
observed properties, it fails in certain aspects, including the
prediction that the system will slowly brighten over the quiescent
phase, which observations do not show. See \citet{Hameury2020} for a
recent review of how predictions of the DIM compares with the
observational proprties of outbursts from compact binaries, including
AM CVn's.

An even earlier model for explaining outbursts from CVs is the Mass
Transfer Instability Model (MTIM) which predicted that outbursts were
due to episodes of increased mass transfer from the donor star
leading to a disc with high viscosity
(\citet{Bath1975,BathPringle1981}). In the last twenty years, evidence
for MTIM in some outbursting systems has increased (see
\citet{Baptista2022} for some examples, including the AM CVn system YZ
LMi). With the increased number of systems over a range of orbital
period, AM CVn binaries can be used to confront both the DIM and MTIM
models with observational evidence.

\section{How to identify AM CVn binaries}

If we aim to understand the long term behaviour of AM CVn's as a
function of $P_{\rm orb}$ (and hence test accretion models) we have
to understand the biases which are present in their discovery.

At the very shortest period, HM Cnc (5.35 min) and V407 Vul (9.48 min)
were discovered using {\sl ROSAT} X-ray observations since they are
strong soft X-ray emitters. These are now thought to be `direct
accretors' where the accretion flow impacts the photosphere of the
more massive white dwarf directly without the formation of any
accretion disc. eRASSU J0608-7040 (6.2 min) discovered in 2024 and
3XMM J0510-6703 (with a longer period of 23.6 min) discovered in 2017
make up their class. Since they do not have accretion discs, we will
not discuss them further.

Systems with periods close to ten minute include ES Cet (10.3 min)
which was discovered in a UV survey and were subjected to high time
resolution photometry by Brian Warner and others which revealed their
strong periodic modulation which is due to super-humps in a disc which
is always in the hot state (and analogous to the nova-like hydrogen
accreting systems). Observations made using ZTF revealed candidates
similar to ES Cet with dedicated high cadence observations made using
HiPERCAM and other instruments revealing ZTF J0545+3843 (7.95
min). This indicates that wide-field surveys can identify very short
period binaries, but followup observations on $>$2 m telescopes are
required to reveal their hydrogen deficient nature and provide data
which have a sufficient cadence to model multi-colour light curves.

\begin{figure}
    \centering
    \includegraphics[width = 0.9\textwidth]{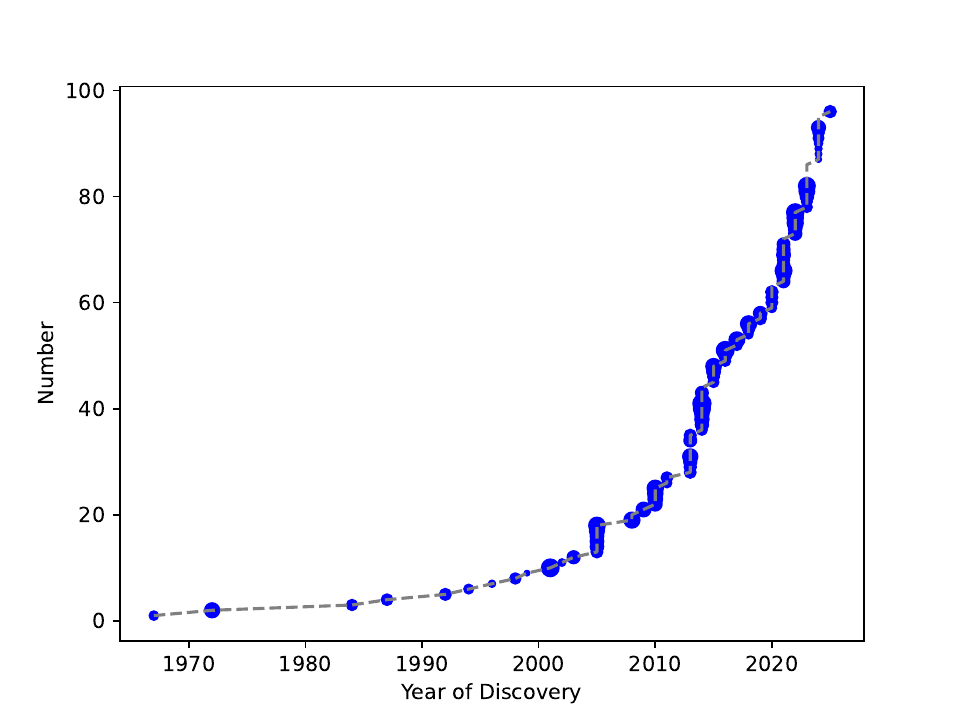}
    \caption{The number of confirmed AM CVn binaries as a function of
      year where the size of the symbol reflects the stars orbital or
      superhump period (data taken from \citet{Green2025}) For systems
      with no confirmed or predicted period we have set their symbol
      size to zero.}
    \label{discovery-year}
\end{figure}

Using the catalogue of \citet{Green2025} we show in Figure
\ref{discovery-year} the number of AM CVn binaries since the discovery
of AM CVn itself in 1967 -- by 1990 only four systems were known,
rising to 25 by 2010 at the point of the \citet{Solheim2010} review.
Clear increases in numbers are apparent in 2005 when the SDSS survey
identified stars with helium lines and no hydrogen. A sharp rise was
also seen in 2013-2014 when transient surveys such as PTF, CRTS and
ASASSN, started to identify outbursts from previously unknown binary
systems. From 2020, ZTF started to make a series of new discoveries of
systems which were outbursting. Because of the sheer number of new
candidate galactic transients from surveys such as ZTF, Atlas,
Panstarrs and GOTO (to name only a few) there is a serious problem of
obtaining spectroscopic data to confirm whether an outburst lacks
hydrogen in its spectrum.  There are 96 confirmed AM CVn binaries in
the catalogue of \citet{Green2025}. Given there are two orders of
magnitude more predicted AM CVn's than are currently known, it is
unclear if many more systems await discovery or that the models which
predict their number are seriously incorrect (see \citet{Rodriguez2025}
for a recent study).

\section{The outburst period range} 

Comparing the long term optical behaviour of AM CVn's with the
hydrogen accreting CVs has long been of interest, with the aim being
to understand what was causing some systems to appear relatively
stable and some which showed outbursts similar to those seen in
hydrogen accreting dwarf novae. Work by \citet{Smak1983} (when there
was two known systems) and \citet{TsugawaOsaki1997} (six systems)
indicated that for systems with $P_{\rm orb}$ \ltae 20 min, the mass
transfer rate was high and the accretion disc would always be in a hot
state. For systems with $P_{\rm orb}$ \gtae 40 min, the disc would
always be in a cool state and would not show outbursts.

By the time of the study of \citet{Ramsay2012a}, who had a long term
programme on the Liverpool Telescope to monitor AM CVn systems, there
were 27 known systems. These authors found that systems with
$P_{\rm orb}$=20-45 min showed evidence for outbursts. For instance, KL
Dra ($P_{\rm orb}$= 24.5 min) shows outbursts every few months. Using the
catalogue of \citet{Green2025} we can reassess whether this finding
still holds.

\begin{figure}
    \centering
    \includegraphics[width = 0.9\textwidth]{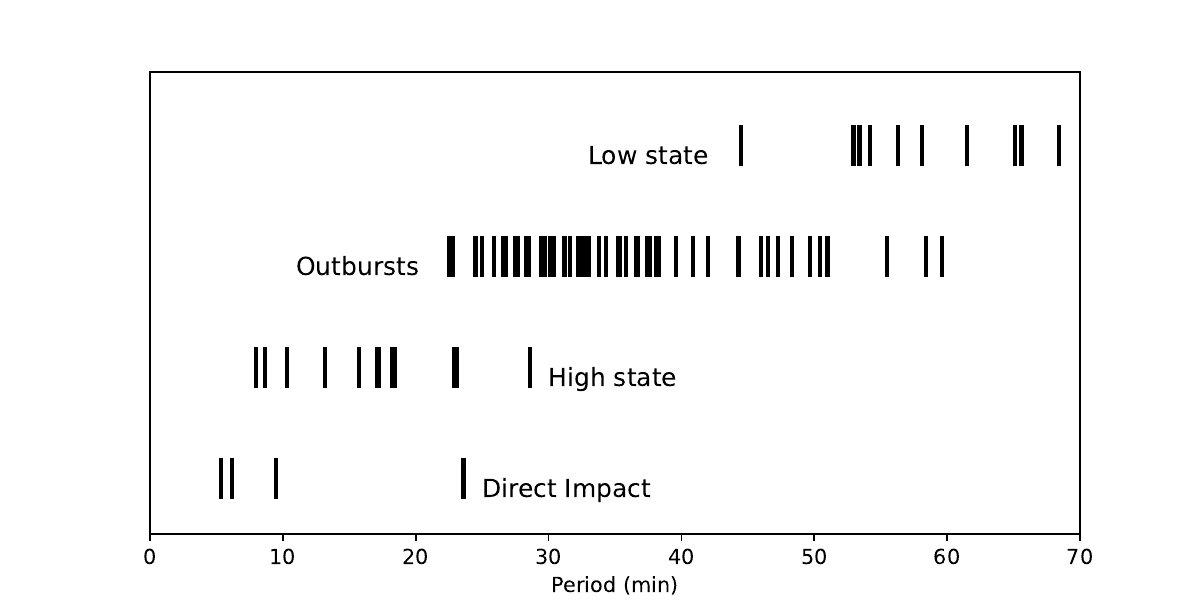}
    \caption{The confirmed systems in the catalogue of
      \citet{Green2025} split into direct imapact, high state,
      outbursters and low state systems.}
    \label{period-state}
\end{figure}

Using $P_{\rm orb}$ (or predicted $P_{\rm orb}$ using the \citet{Levitan2015}
relationship between recurrence time and $P_{\rm orb}$) in
\citet{Green2025} we split the sources into direct impact accretors;
high state systems; outbursting systems and low state systems as
defined in the catalogue and show the results in Figure
\ref{period-state}. As expected with having significantly more
systems, the separation between outbursting systems is not as clear
cut as found in \citet{Ramsay2012a}. (We class GP Com as an
outbursting system since \citet{Kojiguchi2025} and a poster
presentation in these proceedings found an outburst in Harvard plate
scans).

The longer period high state systems are CX361 (22.9 min), which has a
Galactic latitude of -1.4$^{\circ}$, could be less well observed in
wide-field surveys which avoid the plane; TIC 378898110 (23.0 min)
(\citet{Green2024}) did show a long duration and slow increase of
$\sim$0.3 mag but no clear signs of typical accretion outbursts and
SDSS J1831+4202 (23.1 min) which has a '?' beside its high state
status in \citet{Green2025}. All three systems are very worthly of
continued followup to search for future outbursts or unusual
behaviour.

Those systems showing outbursts now extend from ASASSN-14cc (22.5 min)
to SDSS J1137+4054 (59.6 min) with a tendancy for the recurrence
timescale to increase for longer period systems (we will discuss this
later). However, \citet{Duffy2021} used photometry from five
wide-field surveys to study the long term behaviour of eight systems
in a very narrow period range (22.5 and 26.8 min) and found a diverse
set of behaviour implying factors such as formation route, nature of
the donor star, metalicity and potentially the magnetisim of the
accreting white dwarf, play an important role in defining their long
term properties.  Systems in a low state stretch from SDSS J0804+1616
(44.5 min) to the longest period AM CVn system we currently know, SDSS
J1505+0659 (68.4 min). Searching for (possibly) rare outbursts from
these low state systems is strongly encouraged.

\section{KL Dra - a case study of an outbursting system}

\begin{figure}
    \centering
    \includegraphics[width = 0.9\textwidth]{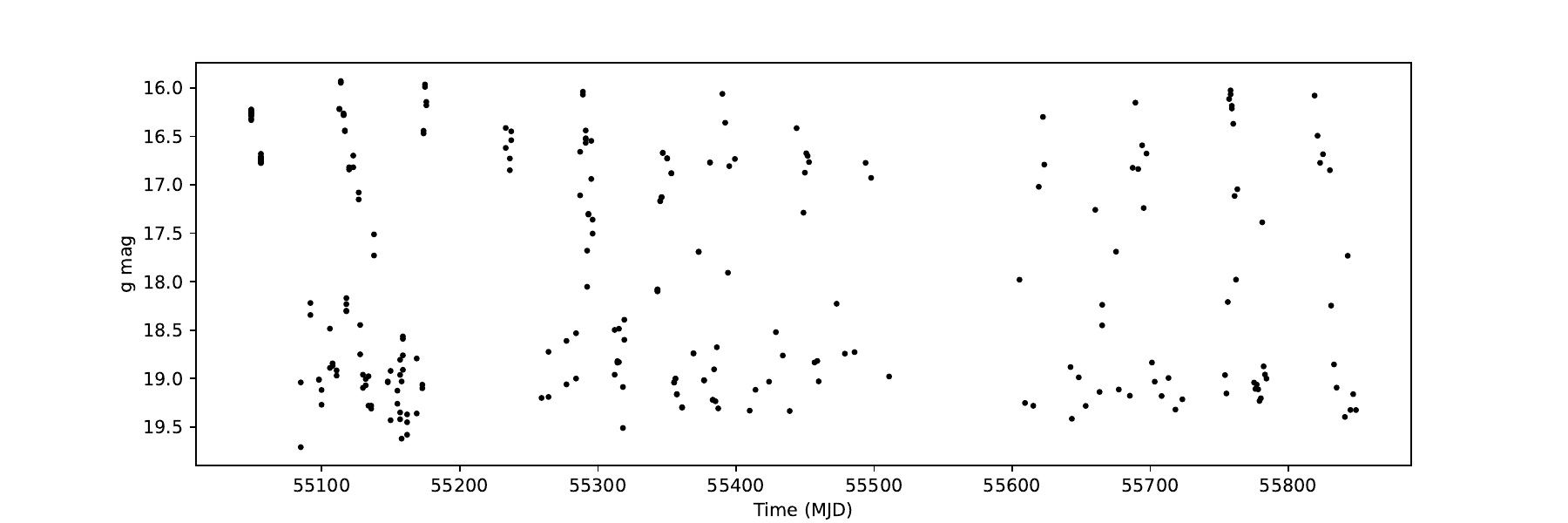}
    \caption{Photometry of KL Dra obtained using the Liverpool
      Telescope between Aug 2009 and July 2011 (data from
      \citet{Ramsay2012a}) showing outbursts every few months.}
    \label{kldra-lt}
\end{figure}

KL Dra was originally identified as a supernova (SN 1998di), although
a subsequent spectrum indicated it was not one, but likely a member of
the (then) small group of AM CVn binaries. Photometric observations by
\citet{Wood2002} revealed a period of 25.5 min (super-humps) in a high
state and 25.0 min (the likely $P_{\rm orb}$) in the low state. In
2009 \citet{Ramsay2012a} started a several year photometric study of AM
CVn's with the Liverpool Telescope on La Palma. It quickly became
clear that KL Dra showed an outburst every few months (Figure
\ref{kldra-lt}), indicating it was an excellent source to study the
accretion process in these hydrogen deficient accreting systems.

With the advent of wide field surveys such as PTF, studies such as
\citet{Levitan2015} were able to identify that some bursts were of
longer duration than others -- i.e. AM CVn's showed normal outbursts and
superoutbursts which are seen in various types of hydrogen accreting
CVs. One of the many achievements of the {\sl Kepler} mission was to
show that superoutbursts in CVs were preceded by a normal outburst
with only a small decrease in flux before the start of the
superoutburst (e.g. \citet{Cannizzo2010}). It was not until {\sl TESS}
observations were made of AM CVn outbursts
(\citet{Duffy2021,PichardoMarcano2021}), that it was revealed that other
systems also show a normal outburst just before a superoutburst.

\begin{figure}
    \centering
    \includegraphics[width = 0.9\textwidth]{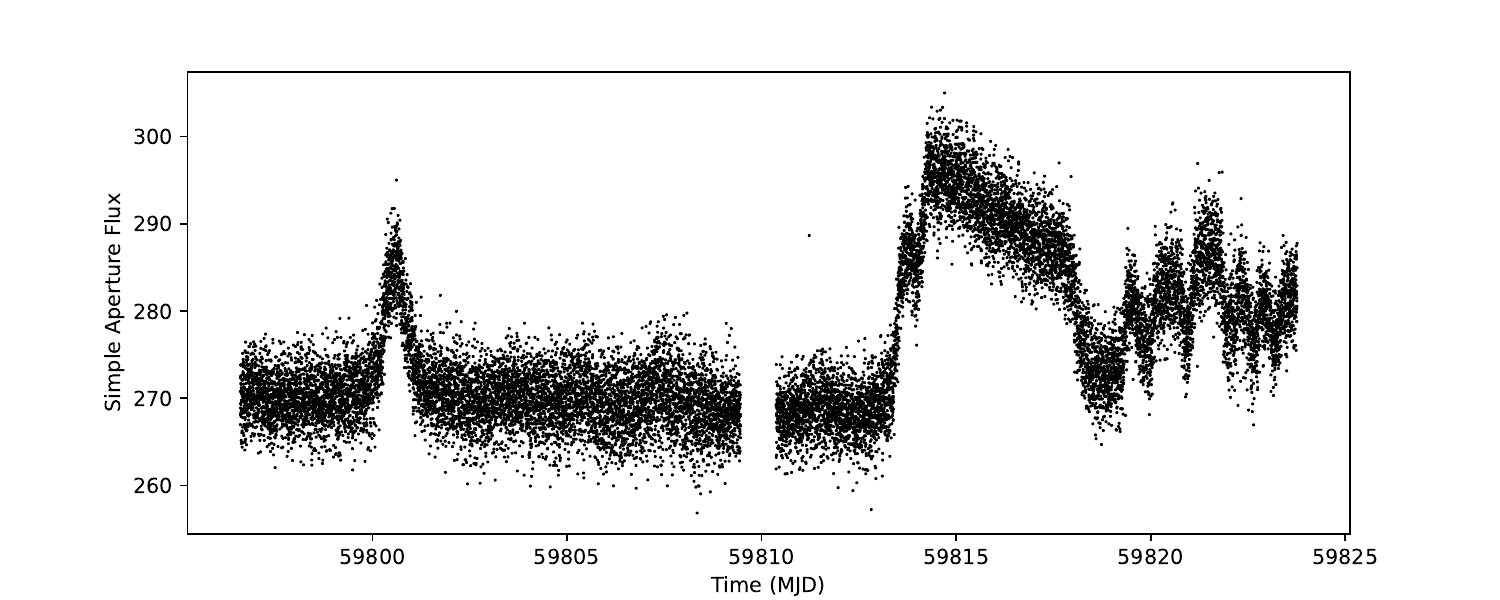}
    \caption{{\sl TESS} data of KL Dra obtained in sector 55 showing one
      normal outburst and one superoutburst (which immediately follows
      a second normal outburst) obtained from the {\sl TESS} SPOC pipeline
      (\citet{Caldwell2020}).}
    \label{kldra-tess}
\end{figure}

{\sl TESS} has now observed KL Dra in dozens of sectors (it is close
to the northern ecliptic pole) and we show the light curve obtained in
one sector using the {\sl TESS} SPOC pipeline in Figure \ref{kldra-tess} and
shows one normal outburst and a superoutburst. The combined
sector-by-sector data show a dozen superoutbursts and three dozen
normal outbursts and are a rich resource for studying the accretion
process in hydrogen deficient binaries.

\begin{figure}
    \centering
    \includegraphics[width = 0.8\textwidth]{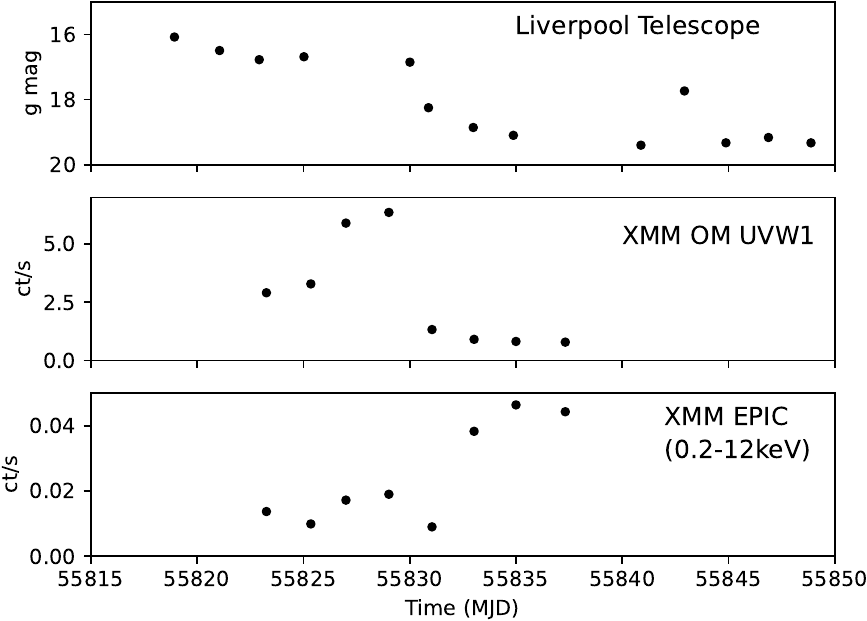}
    \caption{Observations of KL Dra made using the Liverpool Telescope
      (optical); XMM-Newton OM (UV) and XMM-Newton EPIC (X-rays)
      (Adapted from \citet{Ramsay2012b}).}
    \label{kldra-xmm}
\end{figure}

Given we could reasonably predict when an outburst of KL Dra would
take place, we submitted a ToO proposal to observe it using {\sl
  XMM-Newton} in X-rays and the UV and were successful in obtaining
eight pointings over a superoutburst over Sept-Oct 2011. We show the
light curves in the X-ray and UV and also optical data taken over the
same time period using the Liverpool Telescope in Figure \ref{kldra-xmm}.

A superoutburst was ongoing immediately before the start of the
XMM-Newton observations, with the UV flux starting to rise and the
X-ray flux being suppressed. By the end of the superoutburst, the UV
flux dropped suddenly with the X-rays showing a rise. The X-rays
hardened over the course of the bursts, afterwards showing a softening
(\citet{Ramsay2012b}). This is due to the boundary layer between the
accretion disc and the white dwarf becoming optically thick and the
bulk of the emission is released at UV wavelengths.  The same
anti-correlation between X-rays and optical is also seen in an
outburst from SDSS J141118+481257 ($P_{\rm orb}\sim$46 min,
\citet{SandovalMaccarone2019}) while the reverse seems to be found in
ASASSN-21au ($P_{\rm orb}\sim$58 min) during an outburst which showed an
unusually long rise to peak outburst (\citet{RiveraSandoval2022}).

Comparing the characteristics of this KL Dra superoutburst with
hydrogen accreting dwarf nova, we find SU UMa also shows hard X-rays
to be supressed during optical outbursts (\citet{Collins2010}). Hard
X-rays are also suppressed during an outburst from SS Cyg although the
extreme UV emission follows the optical flux (\citet{Wheatley2003}). In
contrast, U Gem shows at least one outburst in which the hard X-rays
follow the behaviour of the optical and UV emission
(\citet{Mattei2000}). It is unclear if these differences are due to a
viewing angle dependance or other effects.

\section{Tests of the disc instability model}

With the arrival of systematic programmes to study the long term
optical behaviour of AM CVn binaries
(e.g. \citet{Ramsay2012a,Levitan2015}), it became possible to
determine if characteristics such as superoutburst recurrence
timescale, duration and amplitude were related to $P_{\rm orb}$.

Using data from 11 systems, \citet{Levitan2015} found there was a
strong correlation between $P_{\rm orb}$ and recurrence time; a slightly
weaker correlation with outburst duration and a lesser correlation
with outburst amplitude. In other words, systems with longer orbital
periods tended to show outbursts separated by longer time intervals
and with longer duration than shorter period systems. 

\subsection{Outburst recurrence time}

\begin{figure}
    \centering
    \includegraphics[width = 0.7\textwidth]{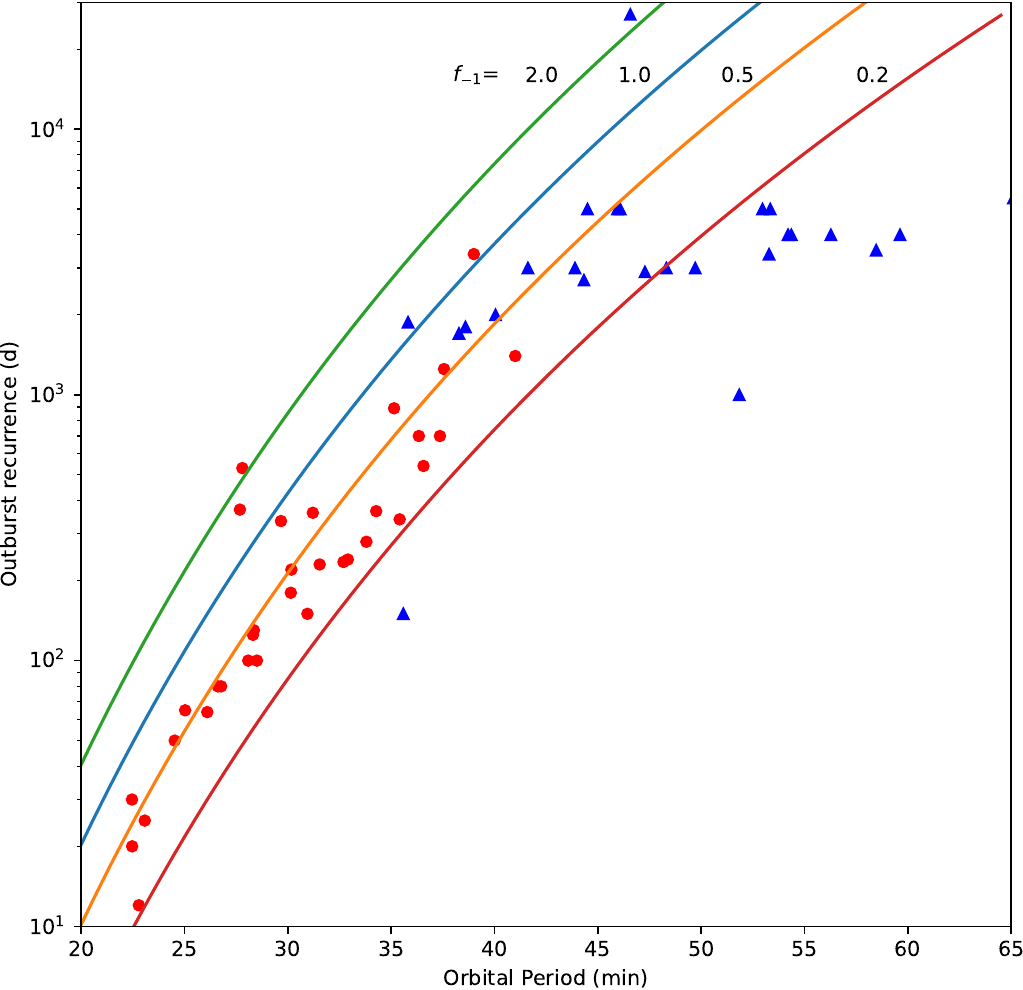}
    \caption{The recurrence time of superoutbursts from the sample of
      \citet{Kojiguchi2025}. We also show the predicted relationship
      using equation 1 (which is equation 18 of
      \citet{CannizzoNelemans2015}) assuming four different values of
      $f_{\rm -1}$.}
    \label{outburst-recurrence}
\end{figure}

\citet{CannizzoNelemans2015} confronted the observational findings of
\citet{Levitan2015} with predictions based on the DIM along with
various assumptions. They found the correlation between the recurrence
time and orbital period was consistent with the predictions of the
DIM. \citet{Kojiguchi2025} collated a much larger sample of recurrence
times of the superoutbursts from 59 systems (some of which were upper
limits) which we reproduce in Figure \ref{outburst-recurrence}. We
also plot the predicted recurrence time using equation 18 of
\citet{CannizzoNelemans2015}:

\begin{equation}
  t_{\rm recur} = (7.59 {\rm days}) f_{\rm -1} \alpha_{\rm c-1}^{-0.82} m_{\rm 1}^{0.85} (1+q)^{1.03} (P_{\rm orb}/1000 {\rm sec})^{7.51}
\end{equation}

where $f_{\rm -1}$ is the fraction of of material in the disc at the end
of the quiescent phase relative to maximum possible; $\alpha_{\rm c-1}$ is
the viscosity paramater, $\alpha$, of the disc in the quiescent state
(both normalised to 0.1); $m_{\rm 1,2}$ is the mass of the primary and
secondary in solar units, $q=m_{\rm 1}/m_{\rm 2}$.

In Figure \ref{outburst-recurrence} we show the prediction from
equation 1 where we assume $m_{\rm 1}$=0.6, $q$=0.05, $\alpha_{\rm
  c-1}$=1, and $f_{\rm -1}$ ranging from 0.2--2.0. We find that most of
the systems are contained in the range between $f_{\rm -1}$=0.2--1.0
indicating $f=0.02-0.1$ which is slightly lower than the expectation
of $f\sim0.1-0.3$ (\citet{CannizzoNelemans2015}), implying the disc
at the end of the quiescent phase has slightly less mass compared to
the initial expectations of these model assumptions. (Although
\citet{Kojiguchi2025} do not include uncertainties on the recurrence
time, for KL Dra this is $\sim$10\% (\citet{Ramsay2012a})).

\subsection{Outburst duration}

Although \citet{Levitan2015} did not find as strong a correlation
between outburst duration and $P_{\rm orb}$ as for recurrence time, it was
still statistically significant. This is maybe due to the sampling
rate of the observations or the complex profile of the outbursts which
make an accurate determination more uncertain. The duration of the
outburst is related to the viscous timescale:

\begin{equation}
  t_{\rm viscous} = (15.1 {\rm days})  \alpha_{\rm h-1}^{-0.8} m_{\rm 1}^{0.34} (1+q)^{0.16} (P_{\rm orb}/1000 {\rm sec})^{0.36}
\end{equation}

which is equation 21 of \citet{CannizzoNelemans2015} where
$\alpha_{\rm h-1}= \alpha_{\rm hot}/0.1$ and other symbols as before.

\begin{figure}
    \centering
    \includegraphics[width = 0.7\textwidth]{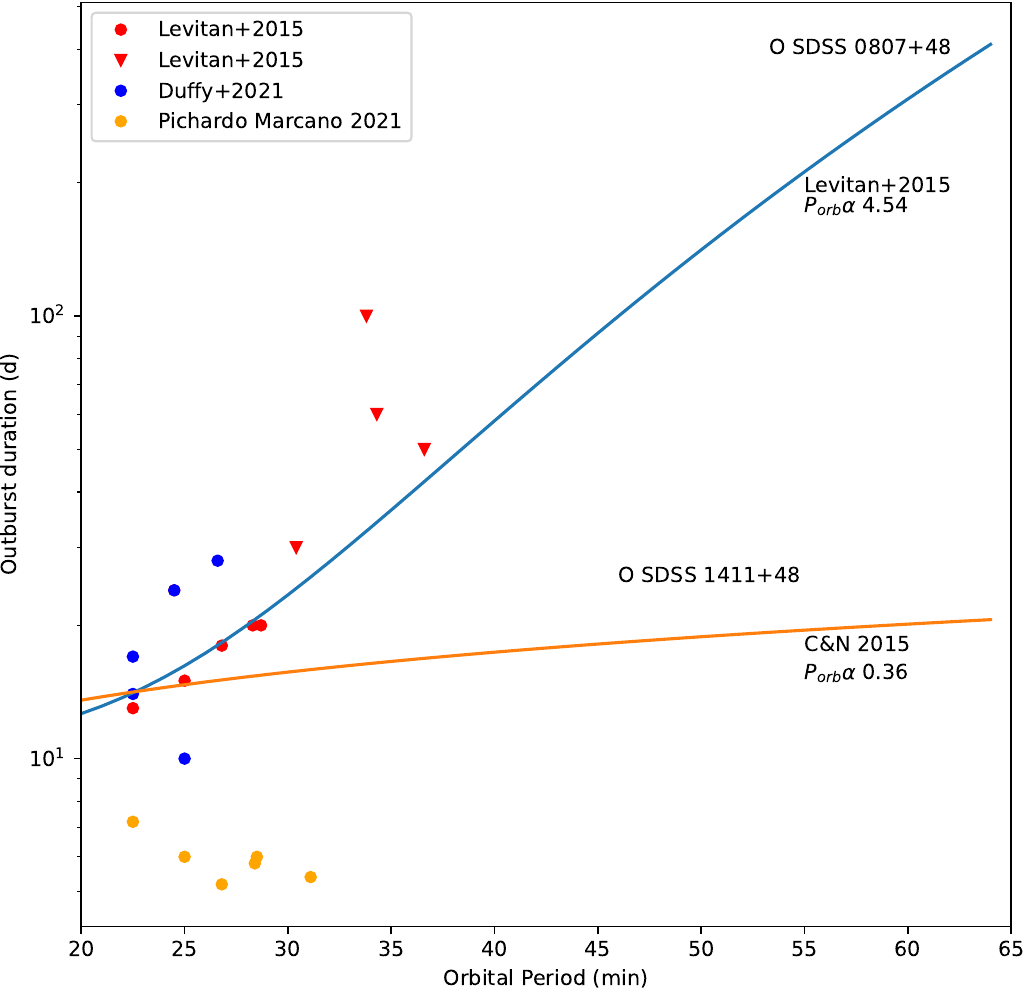}
    \caption{The duration time of superoutbursts and as function of
      $P_{\rm orb}$ taken from \citet{Levitan2015} (including upper
      limits), \citet{Duffy2021} and \citet{PichardoMarcano2021}. We
      also show the initial empirical fit to the data of
      \citet{Levitan2015} and the predicted relationship of
      \citet{CannizzoNelemans2015}).}
    \label{duration-period}
\end{figure}

In Figure \ref{duration-period} we show the duration of outbursts from
\citet{Levitan2015} together with their observed best fit to their
data and also the predicted correlation based on equation 2. We also
add the results from \citet{Duffy2021} who analysed systems in a
similar period range to \citet{Levitan2015}.  Even if we only consider
these samples, which are strongly biased towards shorter period
systems, they do not strongly follow the viscous time relationship.

We now examine two outbursting systems with longer orbital periods:
SDSS 0807+48 (\citet{RiveraSandoval2020}) which showed a very long rise
to outburst ($\sim$200 day) and then an apparent more rapid decline
from maximum and was classed as having a duration of 390 days.  SDSS
1411+48 has shown several outbursts (\citet{SandovalMaccarone2019} plus
AAVSO data) which indicate a much shorter duration. Both of these
systems are shown in Figure \ref{duration-period}: SDSS 0807+48 is
above the empirical \citet{Levitan2015} relationship based on shorter
periods while SDSS 1411+48 is slightly above the viscous relationship
outlined in \citet{CannizzoNelemans2015}. Both are problematic since
the rise to maximum in SDSS 0807+48 is highly unusual (perhaps due to
a different outburst mechanism) and SDSS 1411+48 has a number of echos
or dips in its light curve.

We now return to the shorter period systems, including those
previously examined by \citet{Levitan2015} and \citet{Duffy2021}. Some
of these systems were observed using {\sl TESS}
(\citet{PichardoMarcano2021}) which revealed detail in the light
curves of AM CVn's not seen before, including the fact that
superoutbursts are preceded by a normal
outburst. \citet{PichardoMarcano2021} determined the mean duration of
the superoutbursts in six systems and these are also shown in Figure
\ref{duration-period}: they are shorter by a factor of five compared
to those studies which determine the duration using ground based
all-sky surveys. This maybe explained by the fact that
\citet{PichardoMarcano2021} defined the duration of the outburst which
did not include any dip or echo features that are not always clear in
ground based data.

It is worth noting how the superoutburst was defined in
\citet{CannizzoRamsay2019}. They note the initial rise is set by the
thermal timescale\footnote{The thermal timescale is the time taken for
the disc to change its temperature due to external forces.},
$\tau_{therm}$, and then the slow decay is set by the viscous
timescale\footnote{The viscous timescale is the time taken for
material to move through a disc due to viscosity.}, $\tau_{visc}$,
followed by a faster decay set by
$(\tau_{visc}\tau_{therm})^{1/2}$. They also note `the relaxation back
to the quiescent state can take much longer than what we define to be
duration of the outburst'. Finally the same authors conclude that it
is not clear whether dips or double outbursts should be included in
the outburst duration. However, it should also be noted that the model
assumptions which \citet{CannizzoNelemans2015} used would not have
been able to account for dips and echo outbursts (see \S \ref{echo}).

The definition of the superoutburst duration in AM CVn's need to be
reassessed. For instance, should the duration include the dip phase
which can occur after the peak, and subsequent rebrightening and any
echo outbursts? A systematic study is required which applies a
consistent definition of the superoutburst duration.

\subsection{Outburst amplitude}

Theoretical studies by
\citet{MineshigeOsaki1983,MeyerMeyerH1984,Smak1984} showed that the
observed amplitude of dwarf nova outbursts can only be reproduced in
simulations if the viscosity parameter, $\alpha$, was different by
approximately a factor of ten, during the cold and hot disc states.
(See \citet{Coleman2016,Jordan2024} for more recent simulations).

\begin{figure}
    \centering
    \includegraphics[width = 0.6\textwidth]{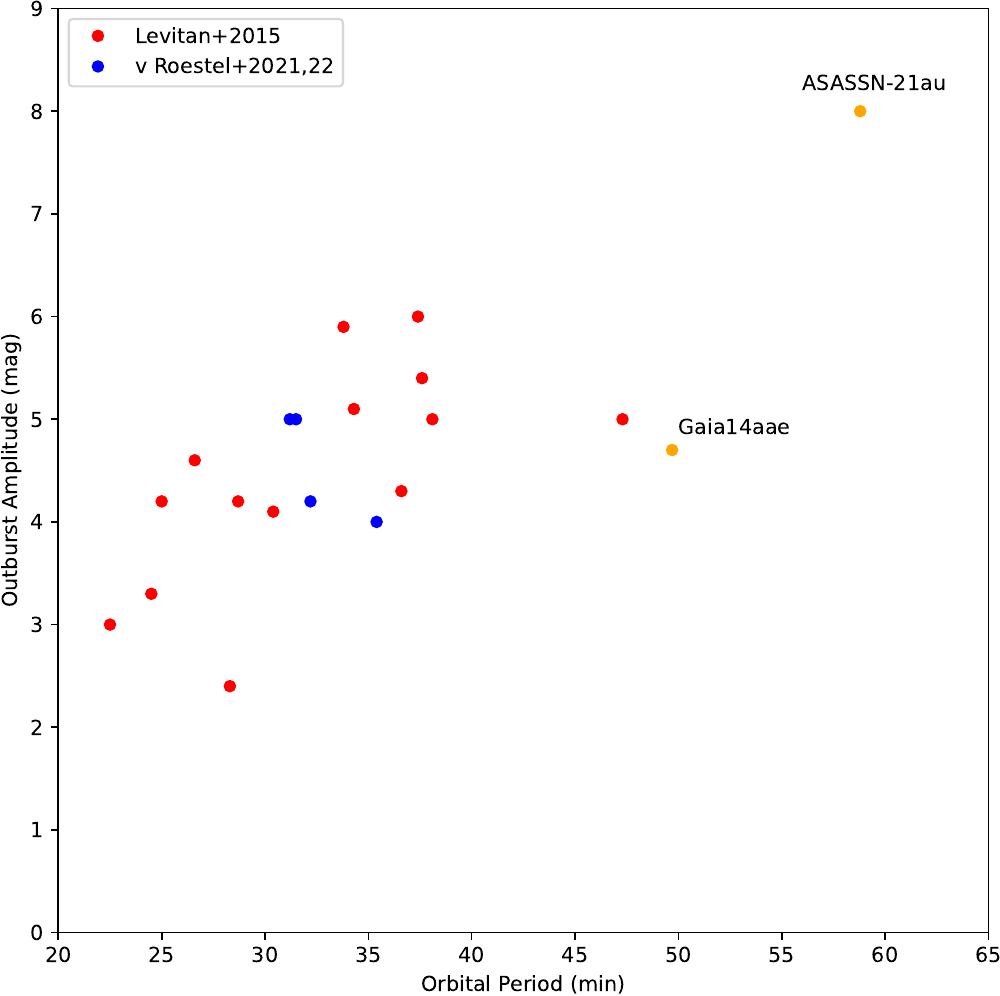}
    \caption{The amplitude of superoutbursts from the samples of
      \citet{Levitan2015,vanRoestel2021,vanRoestel2022} plus two
      longer period systems, Gaia 14aee and ASASSN-21au.}
    \label{amplitude-period}
\end{figure}

\citet{Levitan2015} found evidence of a weak correlation between the
outburst amplitude and $P_{\rm orb}$. Since then many more outbursts have
been observed from many other sources. One of the main issues with
determining the amplitude is the cadence of the observations, which
may miss the peak and that a range of filters are used. Similarly, the
quiescence brightness is not always well established. We show in
Figure \ref{amplitude-period} the amplitude of systems presented in
\citet{Levitan2015} and \citet{vanRoestel2021,vanRoestel2022}. We also
show two longer period systems as examples: Gaia14aee
(\citet{Campbell2015}) and ASASSN-21au (\citet{Isogai2021}). Given that
the distribution of amplitude versus $P_{\rm orb}$ appears far from being
well correlated, a more systematic study of the outbursts of longer
period systems is required.

\subsection{Outburst echos}
\label{echo}

\begin{figure}
    \centering
    \includegraphics[width = 0.9\textwidth]{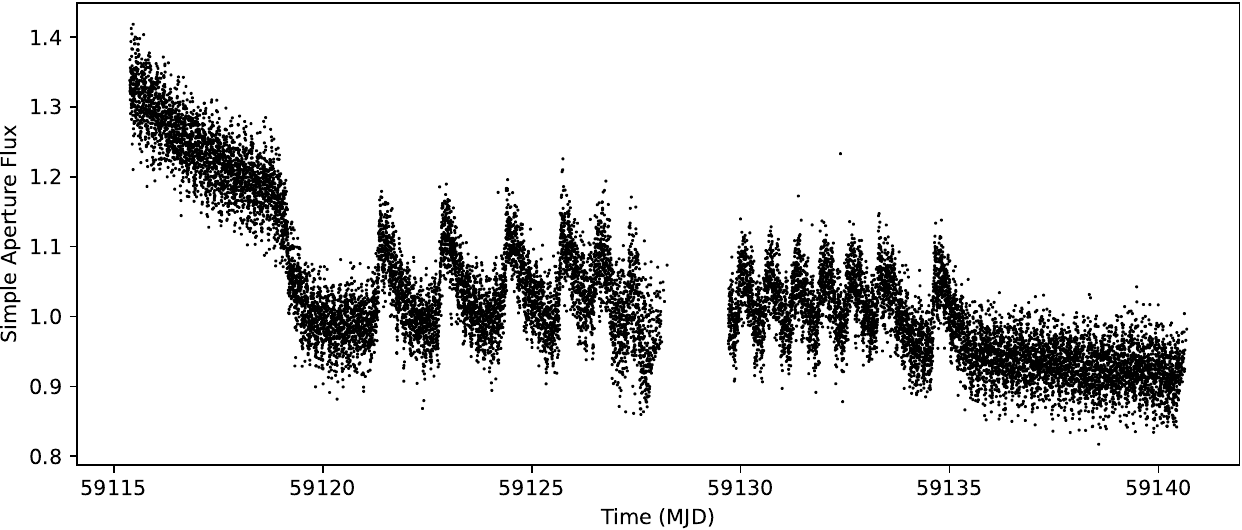}
    \caption{{\sl TESS} observations of Gaia16all made in sector 30 where
      the light curve was extracted using {\sl TESS}-SPOC pipeline (see
      \citet{PichardoMarcano2021} for other observations of this
      source made using {\sl TESS}). This light curve shows a series of echo
      outbursts after the main outburst.}
    \label{echo-gaia16all}
\end{figure}

In some hydrogen accreting CVs, a series of echo, or rebrightenings,
have been seen during the decline from maximum. One famous example is
the EG Cnc 1997 outburst which showed six such bursts
(\citet{Patterson1998}). \citet{HameuryLasota2021} were able to
simulate echo outbursts from several different accreting systems by
incorporating time dependent parameters in the DIM model such as a
varying mass transfer rate and irradiation of the accretion disc from
the hot white dwarf. Several AM CVn systems have also been shown to
show echo bursts, including OX Eri (ASASSN-14ei) and V493 Gem
(ASASSN-14mv) (see \citet{Kato2015} for a review of WZ Sge systems
which also touches on echo outbursts including AM CVn systems). As an
example, we show in Figure \ref{echo-gaia16all} one outburst from
Gaia16all observed using {\sl TESS} which showed a series of echo
outbursts (see \citet{PichardoMarcano2021} for details). These echo
outbursts provide an excellent resource to test the DIM and MTIM
models, including those seen from very compact systems and accretion
flows which are hydrogen deficient.

\subsection{Simulating light curves}

Finally we turn to studies which simulate the light curve of AM CVn's
over multiple outbursts.  \citet{Kotko2010} was able to reproduce
lightcurves of CR Boo and V803 Cen using a constant viscosity
parameter, $\alpha$, and a mass transfer rate close to the upper
critical rate. \citet{Kotko2012} explored this in greater detail and
found enhanced mass transfer was required, likely due to irradiation
of the secondary star, and was able to reproduce the outburst cycle of
KL Dra, $\alpha_{\rm cold}$ = 0.035, using $\alpha_{\rm hot}$ =
0.2. However, it is important to note that normal outbursts were not
confirmed from KL Dra at that point. The fact they are now known to
exist will have an effect on the details of the outburst
model. \citet{Kotko2012} end by saying: `When better and richer sets
of data will become available the comparison of the DIM with
observations will become a precious source of knowledge about
accretion disc physics'.

\section{Conclusions}

With the great increase in the number of AM CVn binaries in recent
years, together with their diversity in outburst properties, their
population is an excellent test for how hydrogen deficient accretion
flows compare with the hydrogen dominated dwarf novae. In particular
the recurrence timescale, duration and amplitude of their outbursts
can be used to test the predictions of the DIM and MTIM models.

Using the recent compliation of \cite{Kojiguchi2025} we find that the
outburst recurrence timescale is quite well described by the
prediction of the DIM, with the caveat that the amount of material in
the disc at the end of the quiescent phase is slightly lower than
originally expected from the model assumptions. When we compare the
duration of the outbursts with the prediction of the DIM and the
empirical relationship of \citet{Levitan2015} we find that different
definitions of the duration make any findings very far from
conclusive. A further study is required which compares the duration of
outbursts using a consistent benchmark. This is also true for
determining the amplitude of outbursts, especially from longer period
systems.

With the diversity and extent of outburst behaviours ranging from
relatively short orbital periods ($\sim$22 min) to long ($\sim$60 min)
there is great scope for the modellers to predict their light curves
using state of the art simulations.

\acknowledgments

I thank the organisers and the Fujihara Foundation for financial
assistance so I could attend the workshop and commend and thank
Matthew Green, Jan van Roestel and Sunny Wong for creating their
excellent catalogue of ultracompact binaries. I also thank the world
wide community of citizen scientists who continue to play a valuable
role in providing the observational material for the study of
accreting binaries in general. This paper includes data collected with
the {\sl TESS} mission, obtained from the MAST data archive at the
Space Telescope Science Institute (STScI). Funding for the TESS
mission is provided by the NASA Explorer Program. STScI is operated by
the Association of Universities for Research in Astronomy, Inc., under
NASA contract NAS 5–26555. I thank the referee for helpful comments.

\end{document}